%% file: main.tex
\documentclass[10pt]{article}

\usepackage[T1]{fontenc}
\usepackage{lmodern}
\usepackage[margin=1in]{geometry}
\usepackage{microtype}
\usepackage{graphicx}
\usepackage{booktabs}
\usepackage{array}
\usepackage{caption}
\usepackage{placeins}
\usepackage{float}
\usepackage{amsmath}
\usepackage{xurl}
\usepackage[numbers,sort&compress]{natbib}
\usepackage{hyperref}
\usepackage{hyperxmp}

\hypersetup{
  unicode=true,
  keeppdfinfo,
  colorlinks=true,
  linkcolor=black,
  citecolor=black,
  urlcolor=black,
  pdftitle={The ack3 H1 2026 DeFi Incident Dataset: Audit Scope Across 135 Security Incidents},
  pdfauthor={Josef Gattermayer, Jan Kalivoda, Arman Ba\v{s}ovi\'c},
  pdfsubject={An empirical study of audit history, audit scope, and audit age across 135 H1 2026 DeFi and crypto-security incidents},
  pdfkeywords={ack3, ack3.ai, DeFi security, Web3 security, blockchain security, smart contract audits, audit scope, audit age, security incidents, security dataset, empirical cybersecurity},
  pdflang={en-US},
  pdfcreator={LaTeX}
}
\newcolumntype{L}[1]{>{\raggedright\arraybackslash}p{#1}}

\input{tables/macros.tex}

\title{The ack3 H1 2026 DeFi Incident Dataset:\\
Audit Scope Across 135 Security Incidents}

\author{
Josef Gattermayer$^{1,2}$ \and
Jan Kalivoda$^{1}$ \and
Arman Ba\v{s}ovi\'c$^{2}$\\[0.5em]
\small $^{1}$ack3 (ACK3 LTD), \href{https://ack3.ai}{ack3.ai}\\
\small $^{2}$Czech Technical University in Prague, Faculty of Information Technology
}
\date{3 August 2026}

\begin{document}
\maketitle

\begin{abstract}
Smart-contract audits cover defined artifacts at a specific time, but the label
\emph{audited} is often treated as project-wide assurance. We analyze audit
history and incident-path scope across \PublishedIncidentRows{} DeFi security
incidents using the H1 2026 DeFi Incident Dataset published by cybersecurity
company ack3\footnote{\url{https://ack3.ai}}, covering 1 January--29
June 2026. The corpus reports
\PublishedTotalLoss{} in attributed loss. We identified public pre-incident
audit history for \PublishedAuditedRows{} incidents: \PublishedOutsideRows{}
incident paths lay outside every identified audit scope, \PublishedInsideRows{}
fell within at least one, and \PublishedInsufficientRows{} could not be resolved
from public evidence. Within this \PublishedAuditedRows{}-incident subset,
outside-scope incidents represented 67.6\% by count and
\PublishedOutsideLossPct{} of reported loss. The
loss-weighted result was concentrated in two large incidents; excluding both reduced the share to
\SensitivityOutsideLossPct{}, while preserving the direction of the result. We
also describe audit age, temporal loss distribution, and affected project
types. The results show that project-level audit history and incident-path
scope are distinct variables.
\end{abstract}

\noindent\textbf{Keywords:} decentralized finance; security incidents;
smart-contract audits; audit scope; empirical security

\section{Introduction}

A DeFi system can include contracts, proxies, bridges, oracles, privileged
roles, front ends, relayers, signers, and cloud services. An audit covers named
artifacts at a stated time. It does not establish security for every component
or later deployment.

This distinction is often lost in the binary label \emph{audited}. Incident
post-mortems reconstruct security incidents and their technical paths. Audit
scope asks a different question: did the incident path fall within the code,
version, and system boundary examined before the incident?

We study that question using the published ack3 H1 2026 DeFi Incident Dataset
\citep{ack3dataset2026}.

\subsection{Research questions}

\begin{description}
  \setlength{\itemsep}{0.1em}
  \setlength{\parsep}{0pt}
  \setlength{\topsep}{0.25em}
  \item[RQ1] What share of incidents had identified pre-incident audit history?
  \item[RQ2] How did their incident paths relate to the identified audit scopes?
  \item[RQ3] How did loss weighting and the two largest incidents affect the
  outside-scope share?
  \item[RQ4] How old was the closest relevant review for inside-scope incidents?
\end{description}

Security guidance treats review and testing as bounded controls. The EEA
EthTrust specification defines review against stated requirements rather than
universal assurance \citep{eea2025ethtrust}. Solidity guidance recommends
modular contracts, fail-safe mechanisms, and peer review
\citep{solidity2026security}. Pre-audit practice ties review to a frozen
codebase, explicit scope, documentation, tests, and static analysis
\citep{kalivoda2023auditprep}.

\section{Data and methods}

\subsection{Corpus}

Candidate discovery combined DefiLlama's hack list with monthly searches of
crypto-security posts on X, protocol and security disclosures, postmortems,
exploit databases, and targeted web queries. Each candidate was checked against
protocol statements, on-chain records, security-firm analysis, and independent
reporting. For projects with audit history, we searched project and auditor
archives and read every located public pre-incident audit report, recording its
date, revision history, scope, reviewed code, and exclusions.

We included incidents whose attacks occurred from 1 January through 29 June
2026 and whose public evidence supported inclusion as \emph{confirmed} or
\emph{likely}. Out-of-window, unconfirmed, false-positive, and duplicate records
were excluded. Records with matching normalized project names and incident dates
within four days were merged, preserving the better-supported record, the larger
loss estimate, and the combined sources and audit records. The corpus includes
DeFi and adjacent crypto-security incidents affecting on- and off-chain
components; the final total is 135.

Confirmed incidents required at least two independent evidence chains. Likely
incidents were included when the balance of evidence supported occurrence but a
material element remained indirect; three likely records have one recorded
evidence chain. Loss is the best-supported public gross estimate at the dataset
cutoff. Three incidents without a numerical estimate remain in counts but not
loss sums.

\begin{figure}[H]
  \centering
  \includegraphics[width=\textwidth,alt={Flow diagram: 135 published incidents lead to 68 incidents with identified audit history, comprising 46 outside-scope, 20 inside-scope, and 2 unresolved incidents.}]{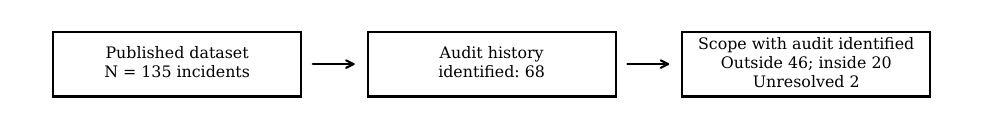}
  \caption{Published dataset and audit-scope derivation.}
  \label{fig:flow}
\end{figure}

Figure~\ref{fig:flow} summarizes the dataset and the 68 incidents with
identified audit history.

\subsection{Variables and analysis}

\emph{Identified audit history} required a public technical-review report or
project audit-archive entry dated before the incident. When a document had
multiple revisions, we used the engagement-end or original final-review date;
later remediation or fix-log appendices did not reset it. \emph{No identified
audit} records a supported negative search; \emph{unknown audit history} records
incomplete evidence.

For each identified review, \emph{confirmed} places the affected artifact or
logic in the reviewed version; \emph{probable} places it in the stated scope
without a revision-level match; \emph{out} places the path outside the review
boundary, including keys, privileged operations, front ends, off-chain
infrastructure, later deployments, or different contracts; and \emph{unknown}
records insufficient evidence.

An incident is \emph{inside scope} if any review is confirmed or probable,
\emph{outside scope} if none is confirmed or probable and at least one is out,
and \emph{unresolved} otherwise. \emph{Likely} grades incident evidence;
\emph{unknown} grades a review; \emph{unresolved} is an incident-level result.

Audit age is the calendar-month difference between the incident and the closest
relevant pre-incident review:

\begin{equation}
A_i=12(y_i-y_a)+(m_i-m_a).
\end{equation}

We use \emph{protocol} as an umbrella term for on-chain systems, including
bridges, lending markets, stablecoins, and decentralized exchanges. The
project-type taxonomy assigns each incident one primary type; \emph{other
protocol} is the residual category for protocols without a more specific type.

We report incident counts and gross reported loss. Scope percentages use only
the 68 incidents with identified audit history; the 35 with no identified
audit and 32 with unknown audit history are excluded. Audit age uses the closest
relevant pre-incident review.
Temporal and project-type views use all 135 incidents. The corpus has no
population denominator. The loss-weighted sensitivity analysis removes the two
largest incidents from its numerator and denominator.

\section{Results}

\subsection{Corpus and audit history}

The published ack3 H1 2026 DeFi Incident Dataset contains
\PublishedIncidentRows{} incidents and \PublishedTotalLoss{} in attributed
loss.

\noindent\textbf{RQ1.} We identified public pre-incident audit history for
\PublishedAuditedRows{} of \PublishedIncidentRows{} incidents (50.4\%); 35 had
no identified audit and 32 had unknown audit history.
Of the 135 incidents, 122 were confirmed and 13 likely. Median reported loss was
\$413,000; the interquartile range was \$134,860--\$2.60 million
(Table~\ref{tab:summary}).

\begin{table}[H]
  \centering
  \caption{Published H1 2026 dataset summary.}
  \label{tab:summary}
  \resizebox{0.72\textwidth}{!}{\input{tables/published_baseline.tex}}
\end{table}

\begin{figure}[H]
  \centering
  \includegraphics[width=0.90\textwidth,alt={Two bar charts. By count: 68 incidents had identified audit history, 35 had no identified audit, and 32 were unknown. Reported loss was USD 721.2 million, USD 109.2 million, and USD 109.4 million, respectively.}]{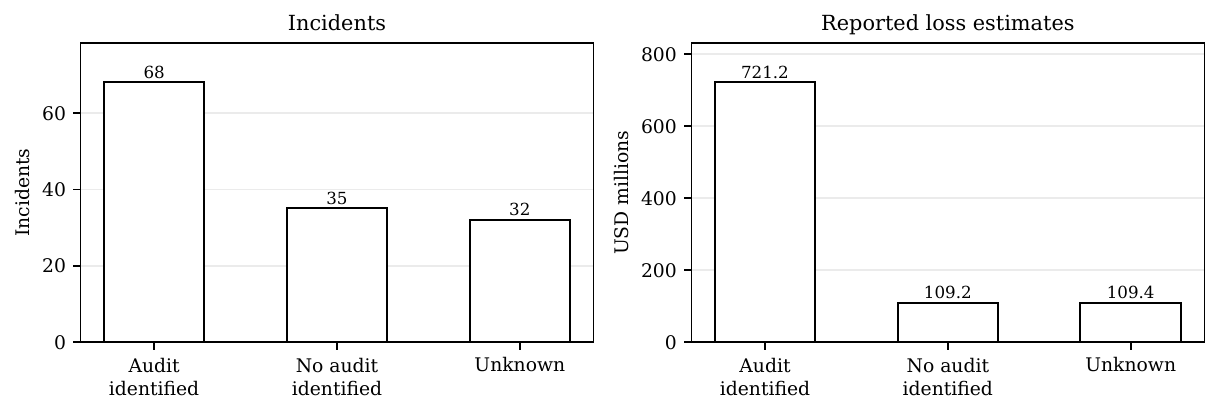}
  \caption{Audit-history status by incident count and reported loss.}
  \label{fig:auditstatus}
\end{figure}

Incidents with identified audit history, no identified audit, and unknown audit
history carried \$721.24 million, \$109.17 million, and \$109.45 million in
reported loss, respectively (Figure~\ref{fig:auditstatus}).

The largest incidents were Kelp DAO (\$292 million; a forged bridge message
after off-chain verifier and RPC compromise) and Drift Protocol (\$285 million;
compromised multisig authority followed by attacker-controlled collateral and
oracle parameters). Truebit Protocol lost \$26.44 million through an unchecked
integer overflow in a legacy bonding-curve contract.

\subsection{Audit scope and age}

\noindent\textbf{RQ2.} Of the \PublishedAuditedRows{} incidents with identified
audit history, \PublishedOutsideRows{} were outside scope,
\PublishedInsideRows{} were inside scope, and \PublishedInsufficientRows{} were
unresolved.

\begin{figure}[H]
  \centering
  \includegraphics[width=0.90\textwidth,alt={Two bar charts for 68 incidents with identified audit history. By count, 46 were outside scope, 20 inside scope, and 2 unresolved. Reported loss was USD 681.0 million outside scope, USD 35.2 million inside scope, and USD 5.1 million unresolved.}]{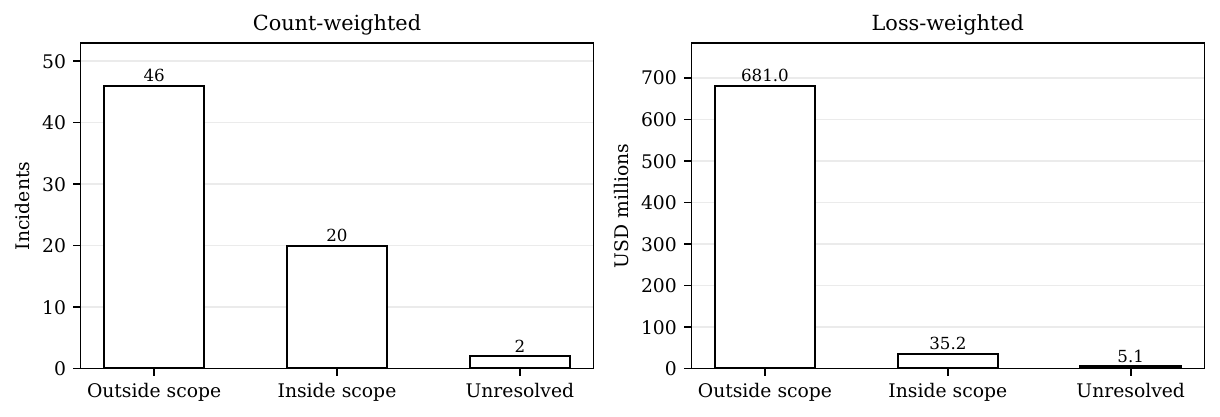}
  \caption{Count- and loss-weighted audit-scope relationships among the 68
  incidents with identified audit history.}
  \label{fig:scope}
\end{figure}

\noindent\textbf{RQ3.} Outside-scope incidents accounted for 67.6\% of the
\PublishedAuditedRows{} incidents by count and \PublishedOutsideLossPct{} of
their reported loss. Reported loss was \$680.97 million outside scope,
\$35.21 million inside scope, and \$5.07 million unresolved
(Figure~\ref{fig:scope}). Excluding the two largest incidents, Kelp DAO
(\$292 million) and Drift Protocol (\$285 million), reduced the outside-scope
loss share to \SensitivityOutsideLossPct{} (\SensitivityOutsideLoss{} of
\SensitivityAuditedLoss{} among 66 incidents).

\begin{figure}[H]
  \centering
  \includegraphics[width=0.58\textwidth,alt={Bar chart of audit age for 20 inside-scope incidents: 3 under six months, 9 from six to under 24 months, and 8 at least 24 months.}]{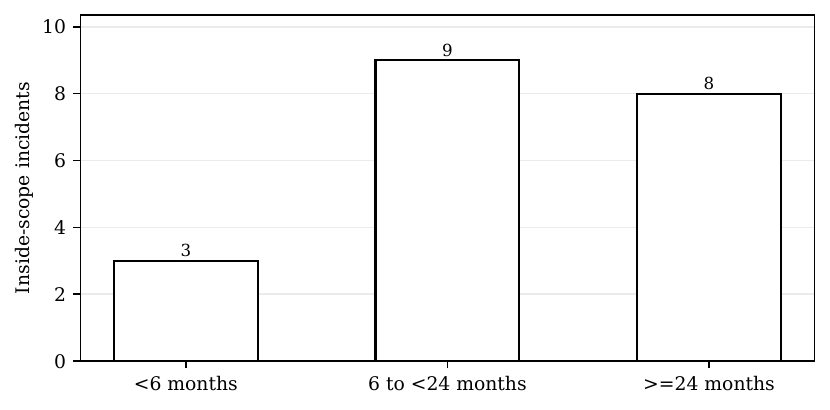}
  \caption{Age of the closest relevant pre-incident review for the 20
  inside-scope incidents.}
  \label{fig:age}
\end{figure}

\noindent\textbf{RQ4.} The 20 observed audit-age gaps ranged from 3 to 56 months,
with a median of 18 months. Three projects were in the under-six-month group:
Makina Finance, Aftermath Finance, and Ekubo Protocol. Reviewed-code hashes were
verified for
Makina (ChainSecurity, \texttt{5bfd17d3}) and Ekubo (Code4rena,
\texttt{2c32cb92}). Nine audits were 6--23 months old, and eight were at least
24 months old (Figure~\ref{fig:age}) \citep{ack3dataset2026}.

\subsection{Temporal and project-type distribution}

\begin{figure}[H]
  \centering
  \includegraphics[width=\textwidth,alt={Scatter plot of incident date versus log-scale reported loss for 135 incidents, grouped by audit-history status. Losses range from no numerical estimate to USD 292 million; incidents with identified audit history include the two largest losses.}]{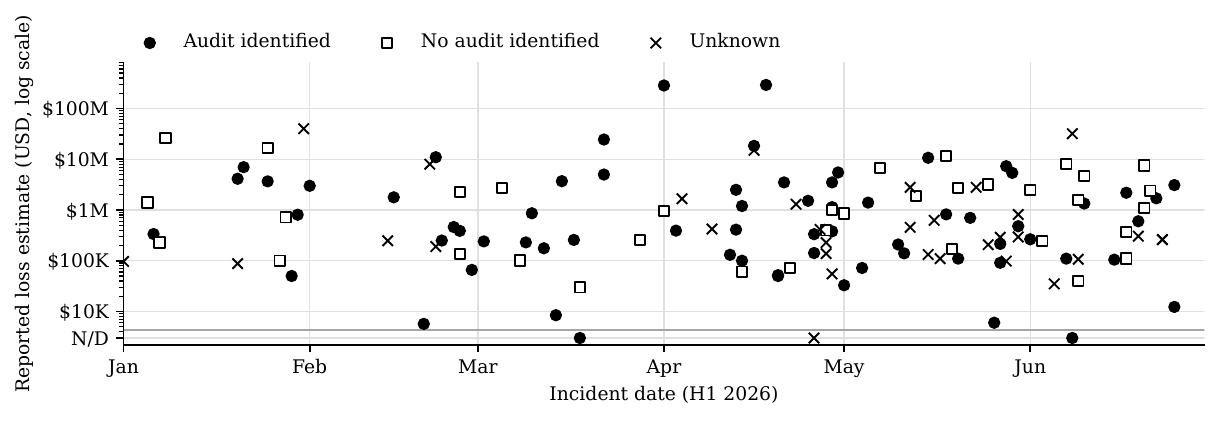}
  \caption{Incident-level reported loss over time by audit-history status.
  Three incidents without numerical loss estimates are shown at N/D.}
  \label{fig:lossdist}
\end{figure}

Incident frequency increased in April and May, while reported losses varied by
several orders of magnitude. Incidents with identified audit history spanned
the observable loss range and included the two largest losses
(Figure~\ref{fig:lossdist}).

\begin{figure}[H]
  \centering
  \includegraphics[width=\textwidth,alt={Two dot plots by affected project type. The residual category other protocol leads with 48 incidents and USD 392.9 million in reported loss. Perpetual-futures exchanges account for 6 incidents and USD 293.7 million; bridges account for 17 incidents and USD 63.2 million.}]{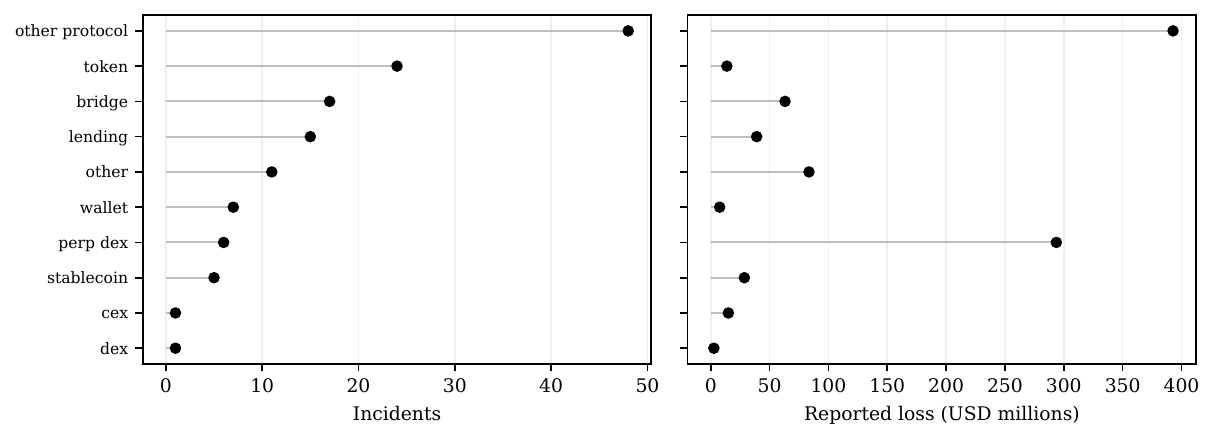}
  \caption{Incident count and reported loss by affected project type.
  \emph{Other protocol} denotes protocol incidents without a more specific type.}
  \label{fig:types}
\end{figure}

The residual category \emph{other protocol} led by count and loss, with 48
incidents and \$392.86 million.
Perpetual-futures exchanges contributed \$293.66 million across six incidents;
bridges accounted for 17 incidents and \$63.15 million
(Figure~\ref{fig:types}).

\FloatBarrier
\section{Discussion}
\enlargethispage{2\baselineskip}

The central finding concerns review boundaries, not audit effectiveness. Among
the 68 incidents with identified audit history, 46 had an incident path outside
all identified pre-incident audit scopes. Taiko Bridge involved an off-chain
prover signing key; Polymarket,
a third-party front-end script; and Resolv / USR, a credential path from GitHub
and cloud infrastructure to on-chain minting authority
\citep{ack3dataset2026}. These cases show why project-level audit labels obscure
the component, version, and time actually reviewed.

This distinction has reporting and maintenance implications. Audit reports
should identify repositories, commits, deployed addresses, exclusions,
privileged roles, and dependencies \citep{kalivoda2023auditprep}. Draft ERC-7512
would encode chain, deployment, time, hash, and URI for audits, but remains a
draft \citep{eip7512}. Because code and surrounding systems change, teams should
repeat review after relevant changes. Audit age measures elapsed time, not
effectiveness.

The findings favor methods that examine cross-component paths. Manually guided
fuzzing in the Wake testing framework exercises reviewer-defined flows and
invariants \citep{ackee2026wake,gattermayer2024mgf}; deployment monitoring
compares reviewed and live code. At ack3, a human-supervised AI scan runs in
parallel with manual review, extending beyond the scoped codebase
\citep{ack3aiscan2026}.

\section{Limitations}

The corpus contains reported incidents, not unexploited protocols or exposure
time; it cannot estimate audit effectiveness, incident risk, or causal effects.
Coverage depends on public reporting and excludes private incidents and
undisclosed audits.

Loss estimates vary in valuation and recovery treatment, and large incidents
dominate loss-weighted results. Scope and system-layer annotations were
single-coded from public evidence, so no inter-rater statistic applies. Results
are limited to this half-year and corpus boundary.

\section{Conflict of interest}

Josef Gattermayer and Jan Kalivoda are affiliated with ack3 (ACK3 LTD), a
cybersecurity company that published the dataset and provides manual security
reviews with human-supervised AI scans that run in parallel with manual review,
extending beyond the scoped codebase.

\section{Data availability}

The frozen version 1.0 dataset is archived at
\url{https://doi.org/10.5281/zenodo.21906487}.
The interactive report is available at
\url{https://ack3.ai/research/defi-hacks-h1-2026/}.

\section{Conclusion}

Among 68 published incidents with identified pre-incident audit history, 46
incident paths were outside all identified public pre-incident audit scopes, 20
were inside at least one scope, and two were unresolved. Outside-scope incidents
carried 94.4\% of the reported loss across those 68 incidents. Audit history is
a project-level variable; scope classification relates an incident path to an
artifact, version, and time. Audit assurance is artifact-bound and time-bound.

\renewcommand{\bibfont}{\small}
\setlength{\bibsep}{0.35em}
\bibliographystyle{plainnat}
\bibliography{references}

\end{document}

%% file: tables/macros.tex
\newcommand{\PublishedIncidentRows}{135}
\newcommand{\PublishedTotalLoss}{\$939.86M}
\newcommand{\PublishedAuditedRows}{68}
\newcommand{\PublishedOutsideRows}{46}
\newcommand{\PublishedInsideRows}{20}
\newcommand{\PublishedInsufficientRows}{2}
\newcommand{\PublishedOutsideLossPct}{94.4\%}
\newcommand{\SensitivityOutsideLoss}{\$103.97M}
\newcommand{\SensitivityAuditedLoss}{\$144.24M}
\newcommand{\SensitivityOutsideLossPct}{72.1\%}

%% file: tables/published_baseline.tex
\begin{tabular}{lr}
\toprule
Measure & Value \\
\midrule
Incidents & 135 \\
Study period & 1 January--29 June 2026 \\
Total attributed loss & \$939.86M \\
Median loss; interquartile range & \$413K; \$135K--\$2.60M \\
Confirmed / likely incidents & 122 / 13 \\
Identified audit history / none / unknown & 68 / 35 / 32 \\
Outside / inside / unresolved scope (n=68) & 46 / 20 / 2 \\
Outside-scope share of loss (n=68) & 94.4\% \\
\bottomrule
\end{tabular}